\documentstyle[aps,epsfig,float,twocolumn,prl]{revtex} 
\newcommand{\be}{\begin{equation}}
\newcommand{\ee}{\end{equation}}

\begin{document}
\twocolumn[\hsize\textwidth\columnwidth\hsize\csname @twocolumnfalse\endcsname 
\draft
\title{Statistical Mechanics of Interfering Links}
\author{M. B. Hastings}
\address{
Center for Nonlinear Studies and Theoretical Division, Los Alamos National
Laboratory, Los Alamos, NM 87545, hastings@lanl.gov 
}
\date{September 8, 2004}
\maketitle
\begin{abstract}
We consider the statistical mechanics of interfering transmissions in a
wireless communications protocol.  In this case, a connection between two nodes
requires all other nodes within communication distance of the given two nodes
to remain quiet on the given channel.  This leads to an interesting problem
of dimers on a lattice, with a restriction that no two dimers can overlap
or be nearest neighbors.  We consider both an equilibrium and a non-equilibrium,
``greedy" dynamics for the links; the equilibrium
properties of the model are found to exhibit an interesting spin glass
transition at maximum density on certain lattices, while the greedy
construction is related to the problem of random sequential adsorption.
\vskip2mm
\end{abstract}
]
The study of networks has identified a variety of interesting networks
with different topological properties\cite{net1}, with recent interest
considering dynamical processes such as flows on the networks\cite{net2}.
Wireless communication networks provide a practically important
example of information flow on a network.  In this paper, we consider the
problem of the MAC (media access control) layer, which lies below the routing
layer and allows different nodes access to given frequencies.
This layer must enable a large number of nodes to communicate over a finite
number of frequencies with finite signal range, while optimizing the
latency, throughput, and fairness of the protocol\cite{w1}.
In this paper, we show that even the simplest model of
the MAC layer
gives rise to rich behavior, including glassy and
non-equilibrium dynamics.

Recently, the problem of the MAC
layer in such a protocol has been cast as a matching problem\cite{anil}.
Consider some set of nodes, each with given signal range.  As a simple
approximation, the signal range leads to a directed
network of possible communication links between nodes: there is a link
from node $i$ to node $j$ if node $j$ can receive a transmission from node $i$.
As a further simplification, we consider the case in which the network
is {\it undirected}, so that whenever node $j$ can receive node $i$, then
node $i$ can also receive node $j$.

Suppose there is only one allowed frequency.
In this case, suppose two nodes, $i$ and $j$ establish a communication link.
Then, all nodes which are within range of node $i$ or node $j$ must
remain silent, staying off the frequency in question, to avoid interference.
Thus, it is possible for a given subset of links to be active simultaneously
if and only if, for all nodes $i$, if node $i$ is in an active link then
$i$ has exactly one neighbor $j$ which is also active.  Equivalently,
this is the problem of finding a set of dimers in which nodes in different
dimers are at least a distance of $2$ from each other on the network; the
presence of a dimer indicates an active link between the two nodes.
The problem of multiple frequencies is similar, and leads to a coloring
problem\cite{anil2}.  In Fig.~1, we show an example network and
various possibilities of both
allowed and disallowed sets of communication links.

In this paper, we study the statistical mechanics of this problem.  We
consider two different dynamics to generate the links.  In the first case,
we consider summing over all possible allowed sets of dimers, weighting
the covering by a fugacity for the number of dimers.  By taking the fugacity
to be large, we are thus able to find high density configurations of
dimers on the network.  We solve this problem when the network is a Bethe
lattice.  Interestingly, the results depend on the coordination number of
the Bethe lattice: for coordination number greater than 5, there appears to
be a spin glass transition at a finite fugacity, while for coordination
number less than 5, this transition is {\it fixed} at infinite fugacity.
We also consider the problem on various finite dimensional lattices and
speculate on the phase diagram.  The problem is related to a recently
considered statistical mechanics of dimers in the limit of infinite
repulsion between neighboring dimers\cite{smd}, but we identify the novel
possibility of the spin glass phase in this limit.

\begin{figure}[!t]
\begin{center}
\epsfxsize=2.5in
\epsfbox{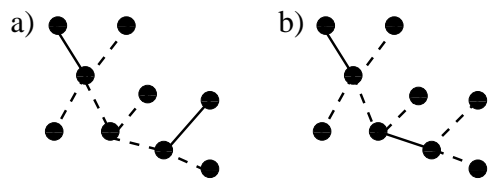}
\end{center}
\caption{a) An example of an allowed configuration of dimers.  Solid
lines denote the presence of a dimer (active link), dashed lines denote
a link on the network which has no dimer (an inactive link).
b) A disallowed configuration of dimers.  The two dimers are within a distance
one of each other.}
\end{figure}

The second dynamics we consider is
a ``greedy" dynamics, in which links are added one at a time to the network,
each link being added at random among the allowed possibilities.  This
dynamics is a version of random sequential adsorption\cite{rsa}.  We
will see that the greedy dynamics does very poorly on graphs with high
coordination number.

To motivate the equilibrium and greedy dynamics, we start with the
following more general set of dynamics: at every instant, we suppose
that some set of dimers is present.  We then suppose that dimers can
be created, when a link is activated, and
disappear, when two nodes which were communicating stop using the channel.
Both processes happen with given rates as follows.
We suppose that at every instant
of time each link which is inactive has some rate of activating (if that does
not interfere with other links), and we set this rate to unity.
We assign
each dimer some rate of disappearing in any instant of time, and we
set this rate to $1/\phi$.
Then, in the stationary distribution of this dynamics, the probability
of finding any given allowed
configuration of dimers is $Z^{-1} \phi^{N_D}$, where $N_D$ is the number of
dimers in the given configuration and
\be
\label{pfn}
Z=\sum\limits_{c} \phi^{N_D}(c),
\ee
where $c$ is a configuration of dimers and $N_D(c)$ is the number of
dimers in the configuration.

Then, the larger $\phi$ is, the larger the average number of active links will
be, thus increasing throughput.  However, a very large $\phi$ means that the
configuration of active links changes slowly in time; the system will tend to
remain stuck in a given configuration for a long time, which is also
undesirable.  We consider two cases below.  First, the stationary dynamics
at finite $\phi$.  Next, we consider $\phi=\infty$, and start from an
initial configuration with no links active; in this case, links can be
activated but are never deactivated.  This leads to a random sequential
adsorption problem, and we will see that the final density of links in this
case is far below the maximum possible in finite $\phi$.

{\it Equilibrium Dynamics on the Bethe Lattice---}
We start with the equilibrium case, solving for the partition function
(\ref{pfn}).  This problem is equivalent to that studied in \cite{smd} in
the limit $\alpha\rightarrow-\infty$ (in that paper's notation).  However,
we consider the possibility of different spin glass and ordering transitions
in this paper.

We consider a Bethe lattice with fixed branching ratio, $k$, so that
every node has $q=k+1$ neighbors.  There are
a number of different possible boundary conditions considered in the 
literature\cite{mz}.
One possibility is to assume that the lattice indeed is a tree with given
branching ratio and with some number of leaves; in this case, appropriate
boundary conditions must be assigned at the leaves of the tree and one must
consider only spins which are remote from the boundary.  This procedure is
reasonable in unfrustrated systems, and in general for systems outside
of a spin glass phase.  However, frustrated systems in the spin glass
phase are sensitive to the boundary
conditions which fix the degree of frustration\cite{lg}.  Then,
a better choice may be a random graph with fixed connectivity $k+1$.

We will solve the problem with a set of recurrence equations on
the Bethe lattice, and identify the onset of a spin glass phase.  In
the spin glass phase, which we do not treat, the system becomes
sensitive to boundary conditions.  Outside the
spin glass phase, the simple solution with recurrence equations is adequate
to describe both the case of a tree with random boundary conditions and
the case of a random graph.

To obtain the recurrence equations, we consider a tree with one of three
possible states for the root of the tree (the root has only $k$ neighbors).
First, the root can be connected to one of its daughters via a dimer (we
call this active).
Second, the root can be not connected to any of its daughters, but have at
least
one of its daughters involved in an active link (we refer to this as locked).
Third, the root can be
not connected to any of its daughters, and have none of its daughters active
either (we refer to this as free).  These three possibilities are
shown in Fig.~2(a-c).

\begin{figure}[!t]
\begin{center}
\epsfxsize=2.5in
\epsfbox{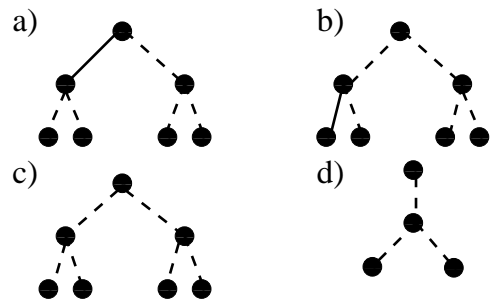}
\end{center}
\caption{a) Active configuration: root connected to daughter by an active
link.  b) Locked configuration: daughter involved in an active link,
preventing root from forming a link to any other node.
c) Free configuration: none of the daughters have active links.
d) Joining $k+1$ trees as described in text.}
\end{figure}

Let the partition function of the dimer problem on a tree in each of the
three cases be $Z_1, Z_2, Z_3$ respectively.  Then take $k$ of these trees
and join them together to make a tree of one higher level, and compute the
resulting partition functions $\tilde Z_1, \tilde Z_2, \tilde Z_3$ for
the larger tree with the three different boundary conditions for that tree.

We have that $\tilde Z_1=\phi k Z_3 (Z_2+Z_3)^{k-1}$,
as one of the $k$ daughters must be
free to have a dimer with the new root, and all the other daughters must
be locked or free (they cannot be active or the root could not make a connection
to the free daughter).
Then, 
$\tilde Z_3=(Z_2+Z_3)^k$,
as each of the $k$ daughters of the new root
must be either locked or free, but cannot be active, so each subtree has
partition function $Z_2+Z_3$.
Finally,
$\tilde Z_2=(Z_1+Z_2+Z_3)^k-(Z_2+Z_3)^k,$
as in this case at least one daughter must be active;
so, we consider all possible states of the daughters, $(Z_1+Z_2+Z_3)^k$, and
subtract off those states with no daughters active, $\tilde Z_2=(Z_2+Z_3)^k$.

It is convenient to rescale the partition functions $Z_i$ by a constant in
looking for a fixed point.  We choose to scale 
$\tilde Z_i\rightarrow \tilde Z_i/\tilde Z_3$,
so $\tilde Z_3=1$.
The rescaled equations become: 
\begin{eqnarray}
\label{req}
\tilde Z_3=1;  \;\;
\tilde Z_1=\phi k (Z_2+1)^{k-1}/(Z_2+1)^k; \\ \nonumber
\tilde Z_2=(Z_1+Z_2+1)^k-(Z_2+1)^k/(Z_2+1)^k.
\end{eqnarray}

We have numerically solved Eqs.~(\ref{req}) and found that for small $\phi$
the solution converges to a fixed point, which we refer to
as a liquid phase.  For larger $\phi$, there is
a range for which the fixed point
is stable, but there is also a stable set of solutions which {\it oscillate}
with period three.  The oscillating solution corresponds to a solid
phase with broken translational symmetry,
in which at one given level of the tree most of the nodes are
free, at the next higher level most of the nodes are active to a daughter,
and at the next higher level most of the nodes are locked.

At even higher values of $\phi$, the stationary solution becomes unstable
and only the solid solution remains.  The existence of the solid
solutions depends on the boundary conditions on the tree.  If the
lattice actually is a balanced tree with given branching ratio
and depth, then the system has a phase transition to a solid phase.
However, on a system on a random graph, the presence of loops will frustrate
the solid phase.  Alternately, on a tree in which the boundary conditions
on the leaves are chosen sufficiently randomly (for example, the tree is not
balanced, but instead unbalanced, so that different leaves are at different
depths below the root), again we can stabilize the liquid phase.
For given $\phi$, there is only one liquid phase (the possibility of
a liquid-gas transition as in \cite{smd} does not happen in this system).

At large $\phi$, we can solve Eq.~(\ref{req}) in the liquid
phase giving
$Z_1^{liquid}=k^{(k+1)/(2k+1)}\phi^{(k+1)/(2k+1)},
Z_2^{liquid}=k^{k/(2k+1)}\phi^{k/(2k+1)}, Z_3^{liquid}=1$.
To find the probability that a given site has a dimer leaving it, we must
then connect $q=k+1$ of these trees into 1 tree, 
as in Fig.~2(d), rather than $k$ trees as above.
After carrying this out, one finds that for large $\phi$ a site has
probability $(k+1)/(2k+1)-{\cal O}(\phi^{-1/(2k+1)})$ of having an active link
emanating from it.  Defining $N$ to be the total number of sites and
$\rho=2N_D/N$, we get a limiting dimer density at large $\phi$ of
$\rho=(k+1)/(2k+1)=q/(2q-1)$.

The limiting density in the liquid phase can be understood on physical
grounds.  For {\it any} lattice with fixed coordination number $k+1$ and with
no loops of length 3, the
dimer density is at most $(k+1)/(2k+1)$: 
consider a pair of sites $1,2$ connected
by a dimer.  Each of those two sites connects to $k$ other sites, giving a total
of $2k$ other sites connecting to the given two sites $1$ and $2$ (by
assuming the absence of loops of length 3, the sites neighboring site $1$
are distinct from those neighboring site $2$).  Thus, for each dimer, we
can identify $2k$ sites which neighbor the given dimer; the total number of
sites which neighbor dimers is then at least $2k/(k+1)$: since each
site has $k+1$ neighbors, we may have overcounted the number of sites which
neighbor dimers by a factor $k+1$.  Then, the total number of sites which
have active links, plus the total number of sites which neighbor a site
with an active link is at least $N_D[2+2k/(k+1)]\leq N$.
Thus, $\rho=2 N_D/N\leq
(k+1)/(2k+1)$.  To achieve this density, it is necessary that if a node
does not have an active link then all $k+1$ of its neighbors have
active links as in Fig.~3(a).
While it appears that the system does achieve this maximum
density as $\phi\rightarrow\infty$
from the results above, we now consider the possibility of a spin
glass transition in the system which may prevent it from reaching the
maximum density.

\begin{figure}[!t]
\begin{center}
\epsfxsize=3.5in
\epsfbox{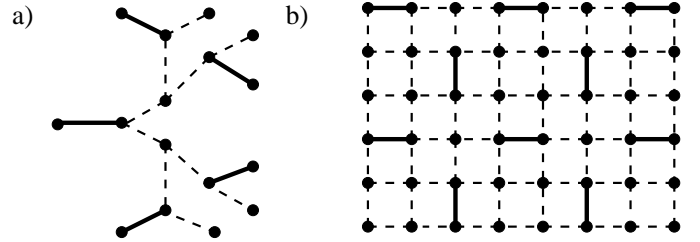}
\end{center}
\caption{a)Configuration to obtain highest density on Bethe lattice.
b) Highest possible density configuration on the square lattice.}
\end{figure}

To identify the spin glass transition, we follow the procedure
used for the Ising spin glass\cite{ith,mz}.
Let each of the $k$ trees that we join together in the procedure above have
slight random variations in the values of $Z_1,Z_2,Z_3$ about the liquid
solution
above, resulting from randomness in the boundary conditions.
If the liquid solution
is stable against these random fluctuations, then the Bethe lattice and
random graph will behave identically in the thermodynamic limit.
Otherwise, we go to a spin glass phase,
and the complete graph may have different behavior.
Fixing $Z_3=1$ as above, we have
two physical quantities,
$Z_1$ and $Z_2$ for each subtree.  Suppose first that on each subtree
each of these two values has {\it the same} slight fluctuation about the
liquid solution above.  We can compute the change in 
$\tilde Z_1$ and $\tilde Z_2$ to linear order
from Eq.~(\ref{req}).  The resulting linear transformation is found to
have eigenvalues
\be
\lambda^{\pm}=-\frac{k \pm \sqrt(k^2-4k)}{2}
\ee
at $\phi=\infty$.  Let these eigenvalues have
left eigenvectors $v^{\pm}$.
Following \cite{ith}, we consider the mean-square fluctuations
$M^{\pm}\equiv \overline{|\sum_i v^{\pm}_i (Z_i-Z_i^{liquid})|^2}$.  Here,
the overline denotes averaging over the random fluctuations, while
the sum over $i=1,2$ projects $(Z_i-Z_i^{liquid})$ onto the corresponding
right eigenvector.
If on each subtree the fluctuations about the
mean are random, then the mean-square fluctuation at the next higher
level is given by $\tilde M^{\pm}=|\lambda|^2/k M^{\pm}$, where the reduction
by $k$ is caused by averaging $k$ random numbers together.  When
$|\lambda|^2=k$, we have a spin glass transition.
Now, for $k<4$, the two eigenvalues are complex conjugates
with absolute value $\sqrt{k}$ at $\phi=\infty$, so that the spin
glass transition occurs precisely at $\phi=\infty$, with no transition
at $\phi<\infty$.

For $k>4$, we have a spin glass transition at finite $\phi$.  The eigenvalues
are both real in this case, with only one of them greater than $\sqrt{k}$
in absolute value; this eigenvalue is negative.  The presence of only
one such eigenvalue suggests that the phase transition to the spin glass
phase might be in the same universality class as the Ising spin glass on
the random graph.

{\it Equilibrium in Finite Dimensions---}
The treatment above is on the Bethe lattice.  The properties of the system
on finite dimensional systems are also of interest.  On a square lattice, the
system has a number of possible ordered structures.
One can show that the pattern of Fig.~3(b) is the highest possible density.
Thus, we conjecture that as a function of $\phi$ the system has a phase
transition from a liquid state at low $\phi$ to the ordered state of
Fig.~3(b) at large $\phi$, with possible additional transitions to
other ordered phases.  A very interesting question is whether the system may
have a glassy phase at some intermediate $\phi$.

{\it Greedy Dynamics---}
The second problem we consider is the random sequential adsorption case at
$\phi=\infty$.  Our technique follows that of \cite{sm} in a slightly different
case of no interaction between dimers.  We start at time $t=0$ with no
links on the lattice.  We then add allowed links at a unit rate per link.
We consider the probability $E_m(t)$ that after time $t$ a connected
cluster of $m$ sites has $(1)$: no links on any of the $m$ sites and
$(2)$: none of the neighbors of the given $m$ sites has a link.  These
conditions imply that the $m$ sites are free to participate in links as
they are not blocked  by any of their neighbors.  It is interesting that
the probability can be written as $E_m(t)$, independent of the structure
of the connected cluster (there are many possible clusters with $m$ sites).
At $t=0$, we have $E_m(t)=1$ for all $m$.  At later times, one can derive
the following differential equation for $m>0$:
\begin{eqnarray}
\label{de}
\partial_t E_m=-(m-1)E_m-[(k-1)m+2]E_{m+1}- \\ \nonumber
[(k-1)m+2]kE_{m+2}.
\end{eqnarray}
There first term corresponds to the possibility of adding a dimer to
any of the $(m-1)$ bonds that connect two sites, both in the cluster.
The second term corresponds
to the possibility of connecting a dimer to any of the 
$[(k-1)m+2]$ bonds which connect
a site in the cluster to a site outside the cluster, while the last term
corresponds to the possibility that a site which neighbors the cluster
will connect to one of the $k$ sites outside the cluster which neighbor
the given site.

The ansatz $E_m(t)=c(t) \sigma(t)^{m-1}$ solves Eq.~(\ref{de}) with
$\partial_t \sigma=-\sigma-(k-1)\sigma^2-(k-1)k\sigma^3$ and
$\partial_t c=-(k+1)\sigma c-(k+1)k\sigma^2 c$ and initial
conditions $c(0)=\sigma(0)=1$.

Define $\rho(t)=2N_D(t)/N$ to be the fraction of 
sites with a dimer emanating from them.  We find $\partial_t \rho=(k+1) E_2$.

Numerical solution of the differential equation gives $\rho(t\rightarrow\infty)
\approx .313$ for $k=4$.  This is noticeably worse than the limiting value of
$\rho=5/9$ computed in the equilibrium case above.  At larger $k$,
the non-equilibrium dynamics does even worse.  The differential
equations can be solved asymptotically at large $k$, where for $c,\sigma>>
1/k$ the equations simplify to $\partial_t\sigma=-k^2 \sigma^3,\partial_t
c=-k^2\sigma^2 c$.  The result for large $k$ is
\be
\rho(t\rightarrow\infty)
\sim \ln(k)/k.
\ee

{\it Discussion---}
We have studied a simple model inspired by recent developments in wireless
communication.  This model turns out to be a model of interacting dimers
very similar to other models studied in statistical physics.  From the
point of view of statistical physics, one of the most interesting features
of the model is the spin glass transition at $\phi=\infty$ for $k\leq 4$.
In the language of statistical physics, this implies that the transition
happens at zero temperature in the mean-field limit.  This may simplify the
treatment of this transition and may be useful for studying finite dimensional
spin glasses.

The spin glass transition is of interest for practical purposes also.  In
in the liquid phase heuristic algorithms will likely
do a good job of rapidly finding allowed dimer configurations of
given density, but in the glass phase it will be much more difficult to find
good configurations and the system will tend to get stuck in certain
configurations.
Finally, the behavior of this system on regular lattices and
random geometric graphs\cite{rgg} in finite dimensions may be interesting.

{\it Acknowledgements---}
I thank E. Ben-Naim, M. Marathe and V. S. Anil Kumar for useful discussions.
This work was supported by DOE grant W-7405-ENG-36.

\end{document}